\documentclass[
 reprint,
 amsmath,
 amssymb,
 aps,
 prl,
 superscriptaddress
]{revtex4-2}
\usepackage{bm}  % bold math symbol
\usepackage{xcolor}  % text color
\usepackage{graphicx}  % figures
\usepackage{xfrac}  % slanted fractions
\usepackage{siunitx}  % units
\usepackage{multirow}  % multiple rows in tables
\usepackage{booktabs}  % better table rules
\usepackage{hyperref}  % hyperlinks
\hypersetup{colorlinks=true, citecolor=blue, urlcolor=blue, linkcolor=blue}
\usepackage{tabularx}  % table columns that auto adjust width

\newcolumntype{C}{>{\centering\arraybackslash}X}  % centered X column for tabularx
\newcolumntype{L}{>{\raggedright\arraybackslash}X}  % left justified X column for tabularx
\newcolumntype{R}{>{\raggedleft\arraybackslash}X}  % right justified X column for tabularx

\begin{document}
\title{Membrane-less phonon trapping and resolution enhancement in optical microwave kinetic inductance detectors}

\author{Nicholas Zobrist}
\email{nzobrist@physics.ucsb.edu}

\author{W. Hawkins Clay}
\author{Grégoire Coiffard}
\author{Miguel Daal}
\author{Noah Swimmer}
\affiliation{Department of Physics, University of California, Santa Barbara, CA 93106, USA}
\author{Peter Day}
\affiliation{Jet Propulsion Laboratory, California Institute of Technology, Pasadena, California 91125, USA}
\author{Benjamin A. Mazin}
\affiliation{Department of Physics, University of California, Santa Barbara, CA 93106, USA}

\date{\today}

\begin{abstract}
Microwave Kinetic Inductance Detectors (MKIDs) sensitive to light in the ultraviolet to near-infrared wavelengths are superconducting micro-resonators that are capable of measuring photon arrival times to microsecond precision and estimating each photon's energy. The resolving power of non-membrane MKIDs has remained stubbornly around 10 at 1 $\mu$m despite significant improvements in the system noise. Here we show that the resolving power can be roughly doubled with a simple bilayer design without needing to place the device on a membrane, avoiding a significant increase in fabrication complexity. Based on modeling of the phonon propagation, we find that the majority of the improvement comes from the inability of high energy phonons to enter the additional layer due to the lack of available phonon states. 
\end{abstract}
\maketitle

To directly image an exoplanet and spectrally characterize its atmosphere, detectors sensitive to light in the 400 nm to 2.5 $\mu$m wavelength range are crucial. Superconducting sensors are important candidates for this application since the low gap energy, $\Delta$, allows for the detection of individual photons. Compared to semiconducting options, superconducting sensors count photons with essentially no noise. This property helps overcome the challenge of the extremely low light levels received from exoplanets~\cite{Walter2020,Steiger2021}. Additionally, in this range of photon wavelengths there are several important spectral features associated with habitability and life which set the required detector performance~\cite{Rauscher2016}. Recommendations for this application in an integral field spectrograph (IFS) typically specify a resolving power of $R = E / \delta E \sim 100$, where $\delta E$ is the full-width half-max energy resolution of the device at the photon energy, $E$~\cite{LUVOIR2019}. However, for space-based instruments where the earth's atmosphere does not interfere, resolving powers as low as 25 can identify large molecular absorption bands like those of water~\cite{Wang2017}.

Several superconducting detector technologies have been proposed for imaging at these wavelengths. Among them are transition edge sensors~\cite{Niwa2017,Burney2006,Romani2001} and superconducting tunnel junctions~\cite{Verhoeve2006,Martin2006}. However, for an exoplanet IFS, tens of thousands to millions of pixels are required to cover the desired field of view which introduces significant wiring challenges in the cryogenic device environment. MKIDs natively solve this issue and achieve similar spectral resolution. Each sensor is a microwave resonator whose inductance and loss temporarily increase after the absorption of a photon~\cite{Day2003}. Photon energies can then be determined by probing each pixel at its resonance frequency and measuring the size of its photon response. This design allows for the straightforward frequency multiplexing of up to 2,000 pixels per feedline and has enabled the demonstration of arrays of up to 20,000 pixels~\cite{Szypryt2017b,Zobrist2019a}.

The limiting resolving power of optical MKIDs is set by the interaction between broken Cooper pairs, also known as quasiparticles, and phonons in the superconductor. The two systems are coupled, and as the energy down-converts from the initial photon absorption, roughly 41\% of the initial energy will be contained in phonons with energies below $2 \Delta$ at which point they can no longer break new Cooper pairs and be detected~\cite{Kozorezov2000}. The exact amount of energy lost in this manner is statistical and sets the maximum achievable resolving power, called the Fano limit~\cite{Fano1947}.
\begin{equation} \label{eq:Rmax}
    R_\mathrm{Fano} = \frac{1}{2 \sqrt{2 \ln(2)}} \sqrt{\frac{\eta_\mathrm{pb} E}{\Delta F}}
\end{equation}
Here, $\eta_\mathrm{pb}$ and $F$ are the pair breaking efficiency and the Fano factor. We use the standard values 0.59~\cite{Kozorezov2000} and 0.2~\cite{Kurakado1982,Rando1992} respectively for each since they are difficult to measure and vary only weakly across most superconductors~\cite{Kozorezov2000,Fano1947}. For a detector with a superconducting transition at 500 mK, the Fano limit gives a maximum resolving power of roughly 59 at 2.5 $\mu$m and 147 at 400 nm, right in the target range for an exoplanet IFS.

In practice, achieving the Fano limit in real detectors has proven difficult. Phonons escaping into the substrate before they fall below the $2 \Delta$ energy threshold can significantly reduce the observed resolving power~\cite{Kozorezov2008, deVisser2021, Zobrist2019}. We account for this excess loss by introducing an extra phonon loss factor, $J$, into equation~\ref{eq:Rmax}~\cite{Kozorezov2007}.
\begin{equation}
    R_{\mathrm{phonon}} = \frac{1}{2 \sqrt{2 \ln(2)}} \sqrt{\frac{\eta_\mathrm{pb} E}{\Delta (F + J)}}
\end{equation}
After measuring the signal-to-noise contribution to the resolving power, $R_\mathrm{noise}$, the total resolving power is given as $1/R^2 = 1/R_\mathrm{noise}^2 + 1/R_\mathrm{phonon}^2$. In this way, $J$ can be estimated for different materials and geometries assuming that there are no other contributions to $R$. The photon event absorption position along with a non-uniform current density in the inductor can also contribute to a decreased resolving power, but modeling of these effects suggests that they do not contribute significantly for this design as long as $R\lesssim40$~\cite{Zobrist2019}.

The best published MKID resolving powers to-date have been in NbTiN-Al hybrid coplanar waveguide resonators suspended on silicon nitride membranes~\cite{deVisser2021}. Compared to that of a device on a thick substrate, the membrane device had a higher resolving power, corresponding to a decrease in $J$ by a factor of about 8, from 3.1 to 0.38. The much thinner membrane allows partially escaped phonons to be quickly recollected in the sensor before reaching the substrate. The extra chance to downconvert into quasiparticles and be detected increases the average time required for phonons to fully escape, $\tau_\mathrm{esc}$, and results in a higher resolving power. This improvement was shown to be consistent with a simple geometric ray-tracing phonon model, which used the proportionality of $J$ to the ratio between the phonon pair breaking time and the escape time, $\tau_\mathrm{pb} / \tau_\mathrm{esc}$, to evaluate the expected decrease in $J$.

\begin{figure}[t]
    \centering
    \includegraphics[width=\linewidth]{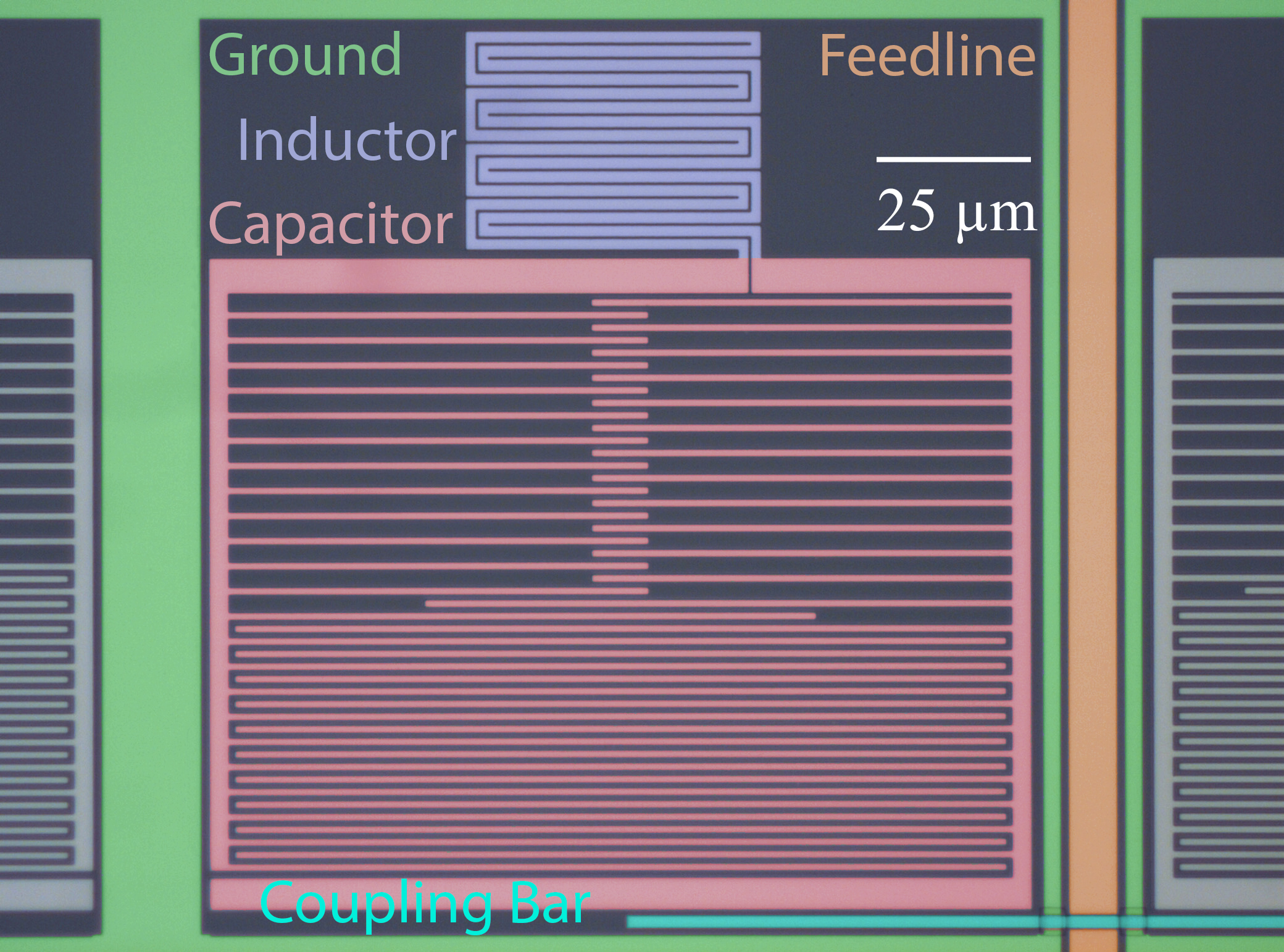}
    \caption{A microscope image of a hafnium optical kinetic inductance detector coupled to a coplanar waveguide feedline. The image has been given false color to highlight the functions of each part of the device. The dark areas are the bare sapphire substrate. Light is focused onto the 47 $\mu$m x 34.7 $\mu$m inductor with a microlens, which allows arrays of these detectors to achieve near unity fill factors. An approximate scale-bar is included for reference.}
    \label{fig:mkid}
\end{figure}

While the membrane devices give an impressive increase in $R$, they introduce significant fabrication complexity. Additionally, the aluminum sensor used in that demonstration is small, highly reflective of optical photons, and has a low kinetic inductance, which makes it difficult to create large arrays with high quantum efficiency. A more realistic detector design for a kilo- or mega-pixel detector requires a more disordered superconductor with a higher kinetic inductance that can be patterned into a compact lumped element circuit like shown in figure~\ref{fig:mkid}. The lumped element design allows for light to be focused onto the inductor with a microlens array making fill factors of $>90\%$ possible. Hafnium has proven to be the best material out of the higher inductance materials tested so far~\cite{Zobrist2019a, Zobrist2020} and has the added benefit of being much less reflective than aluminum~\cite{Coiffard2020, Ehrenreich1963}. We present a measurement of the resolving power of a 220 nm thick hafnium detector on a sapphire substrate with a superconducting transition temperature of $T_c = 395$ mK, in figure~\ref{fig:hafnium}. More details on the resolving power calculation can be found in appendix~\ref{sec:r}. The breakdown of the noise contributions to $R$ shows that athermal phonon escape is the leading factor for the tested energy range, corresponding to $J=13$. This value for $J$ is larger than that for aluminum on silicon nitride, but because of hafnium's lower gap energy, the limiting resolving power is similar.

\begin{figure}[t]
    \centering
    \includegraphics[width=\linewidth]{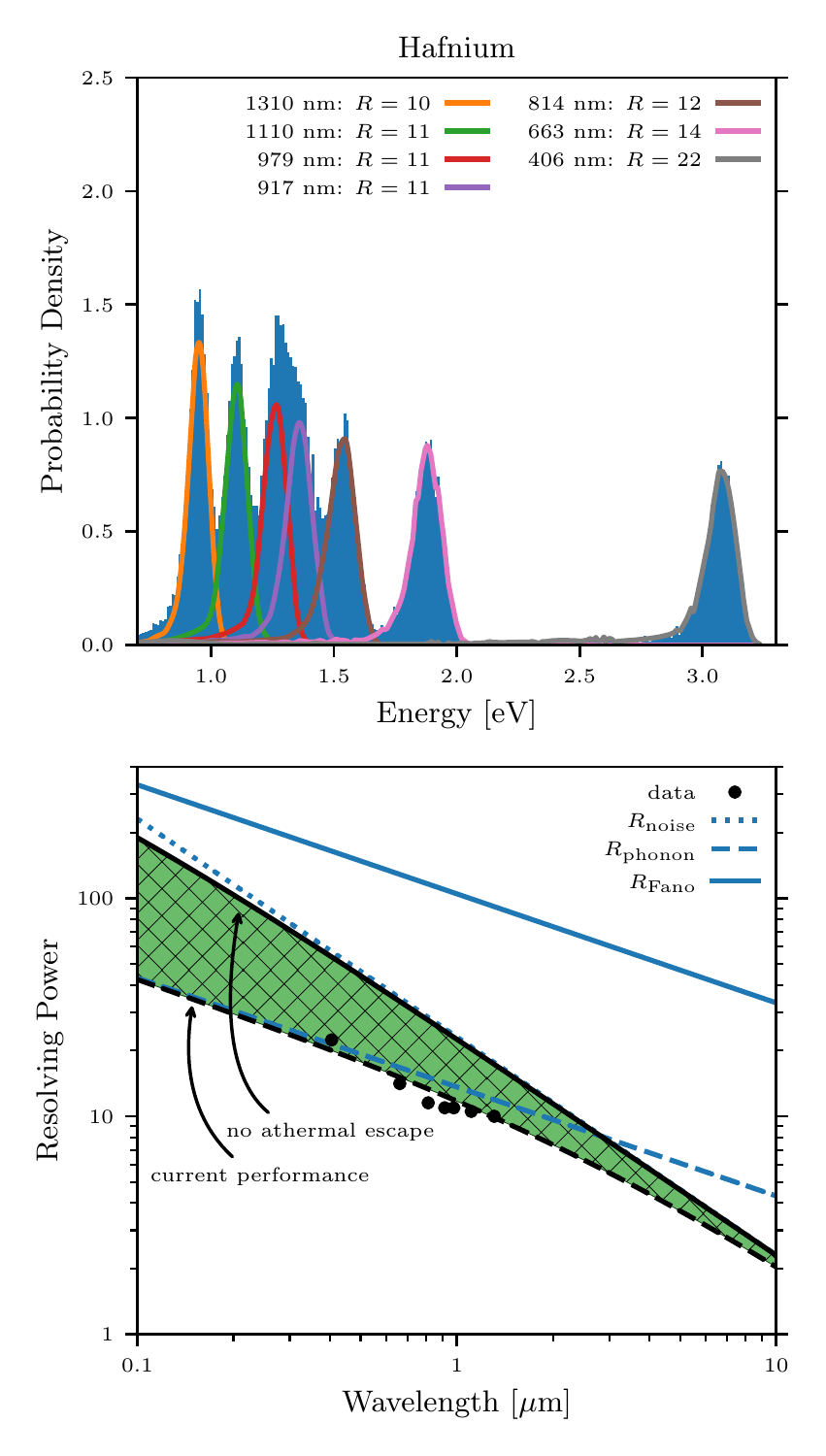}
    \caption{\textit{Top:} Plotted are the combined spectra for the single layer hafnium device at seven laser energies. The small low energy tail to each distribution is likely explained by quasiparticle diffusion into the insensitive capacitor. \textit{Bottom:} The noise decomposition for this device shown. The filled in green area represents the resolving powers achievable by reducing the phonon loss but keeping the same noise spectrum.}
    \label{fig:hafnium}
\end{figure}

To decrease $J$ without a membrane, some form of phonon blocking layer must be introduced between the photo-sensitive superconductor and substrate. This layer may take the form of a material with an acoustic impedance very different from either the substrate or sensor material. In this case, phonons would preferentially be reflected back into the sensor allowing for more opportunities to break Cooper pairs into quasiparticles before falling below the $2 \Delta$ threshold. An example of this kind of layer might be a low density polymer like polymethyl methacrylate (PMMA), which according to the acoustic mismatch model would reduce the effective phonon transmission coefficient into the sapphire by a factor of 3.1~\cite{Little1959,Kaplan1979}. See appendix~\ref{sec:transmission} for more details on the acoustic mismatch calculation.

However, amorphous-insulating blocking layers like PMMA are another potential source of loss and should be kept away from the MKID capacitor to avoid excess two-level system noise~\cite{Vissers2012}. Since most of the escaping phonons contributing to $J$ have energies near the Debye energy of the sensor material, a potential alternative, then, is to find a metallic (preferably superconducting) layer that has a low enough density and speed of sound to not have any available phonon states near that energy. Out of the available elemental superconductors, indium stands out as one of the softest. Indium's effective phonon cutoff energy is much lower than that of hafnium: the Debye temperature of indium, $112$ K, is roughly half of that in hafnium, $252$ K~\cite{McMillan1968}. Unlike with PMMA, adding an indium interface layer would result in only a ${\sim}18\%$ decrease in the effective transmission coefficient according to the acoustic mismatch model. The highest energy phonons produced in the hafnium, though, should be unable to escape into the indium below. These phonons have a wavelength on the order of the hafnium lattice spacing, ${\sim} 0.3$ nm, so indium films that are tens of nanometers thick should provide an effective barrier.

\begin{figure}[t]
    \centering
    \includegraphics[width=\linewidth]{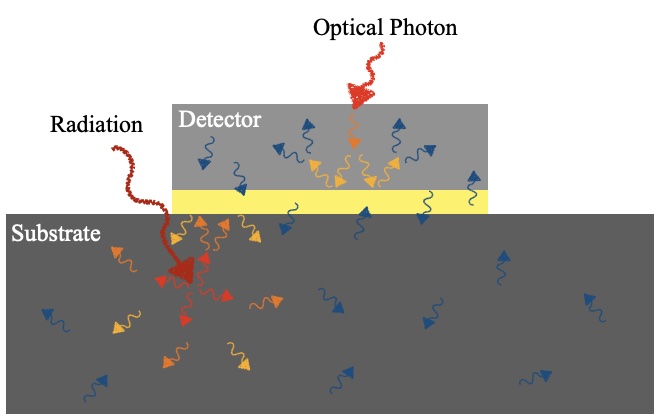}
    \caption{A schematic representation is shown of the phonon blocking layer (yellow) employed in this paper. The lack of available phonon states in the blocking layer prohibits high energy phonons (red and orange) generated during a photon absorption from escaping the detector material, allowing all of their energy to be measured. Lower energy phonons (blue) pass through the barrier freely. Similarly, the phonon blocking layer provides some protection to the detector from high energy events in the substrate from ionizing radiation.}
    \label{fig:diagram}
\end{figure}

Similar types of multilayers with mismatched Debye temperatures have been previously fabricated to produce low thermal conductivity films at room temperature~\cite{Dechaumphai2014}. Because at higher temperatures the contribution of Debye phonons is important, these systems are effective at limiting the quasi-equilibrium heat transfer across the film boundary. Work with these multilayers shows that the simple considerations used here to choose an interface material are likely inadequate for fully describing the phonon transport. The amorphous nature of these films and the presence of the interface, for example, alter the fundamental properties of the phonons~\cite{Giri2020}. However, we will continue to use these simple models as order of magnitude estimates, noting that the future design of this kind of interface layer would benefit from a more detailed analysis.

To demonstrate the phonon blocking effect, we fabricated a hafnium / indium bilayer MKID on silicon, schematically shown in figure~\ref{fig:diagram}. Silicon was chosen as a substrate to encourage more uniform indium films, but we note that thin layers of indium can be deposited on sapphire if the substrate is cooled to liquid nitrogen temperatures~\cite{Chaudhari1965}. The tested bilayer was comprised of a 15 nm layer of indium with 220 nm of hafnium on top. More fabrication details for this device can be found in appendix~\ref{sec:fab}. We measure $T_c = 468$ mK for hafnium on silicon and 786 mK for the bilayer on silicon. Because of the relatively thin indium layer, we expect the film to be proximitized to a single gap energy with a single superconducting transition temperature. However, the increase in $T_c$ is likely mostly due to fabrication differences instead of the proximity effect. See appendix~\ref{sec:proximity} for more details. Resonators patterned out of this material had internal quality factors of up to 250,000, which is similar to the quality factors achieved in hafnium alone~\cite{Coiffard2020}. 

Additionally, we found that the quasiparticle lifetime increased when the indium was added, giving a phase and dissipation lifetime of 404 $\mu$s and 106 $\mu$s for the hafnium device and 452 $\mu$s and 401 $\mu$s for the bilayer. These values represent the maximum quasiparticle lifetime at zero quasiparticle density and were found by fitting the detector response decay to a quasiparticle recombination model~\cite{Fyhrie2018}. More details can be found in appendix~\ref{sec:qp}. The lifetime in the dissipation signal changes the most, which we suspect may be attributable to the higher phonon density inhibiting quasiparticle relaxation into localized, dissipationless states.

\begin{figure}[t]
    \centering
    \includegraphics[width=\linewidth]{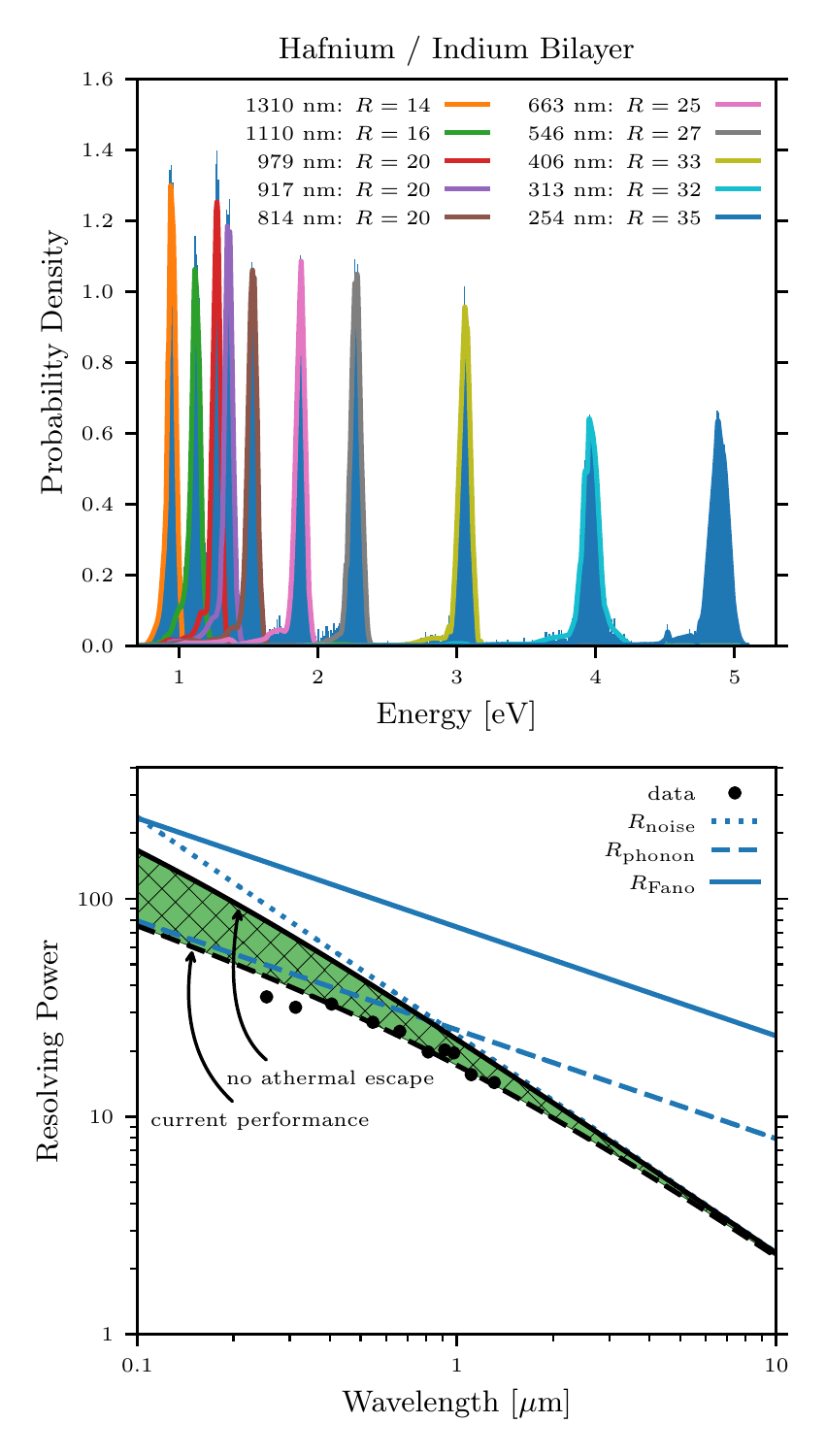}
    \caption{The combined spectra and noise decomposition for the bilayer device are shown, similar to that in figure~\ref{fig:hafnium}. We see a large improvement in the resolving power which is explained by significantly less athermal phonon escape.}
    \label{fig:bilayer}
\end{figure}

Figure~\ref{fig:bilayer} shows a dramatic improvement in the resolving power from 11 to 20 at 1 $\mu$m in the bilayer devices. From this data, we calculate $J=1.6$ corresponding to a 8 times improvement in phonon trapping over the original device, similar to the improvement seen by suspending an MKID on a membrane. The full noise breakdown of this device is also shown in figure~\ref{fig:bilayer} and has been extrapolated to higher and lower wavelengths as a guide to how these detectors are likely to perform outside the tested wavelength range. For the bilayer device above 1 $\mu$m, the resolving power is strongly limited by the signal to noise of the photon pulse, while below 1 $\mu$m phonon escape becomes the limiting term in the resolving power. 

If the increase in energy resolution were from only the acoustic mismatch introduced by the indium layer, we could compute the expected increase in $R$ using the phonon ray-tracing model developed in reference~\onlinecite{deVisser2021}. This model is discussed in more detail in appendix~\ref{sec:sim}. We find that $\tau_\mathrm{esc} = 12$ ns for hafnium on sapphire and $12$ ns for the bilayer on silicon. The ratio of $J$ in the hafnium film to that in the bilayer is given by
\begin{equation}
    \frac{J_\mathrm{Hf}}{J_\mathrm{Bi}} = \frac{\tau_\mathrm{esc, Bi}}{\tau_\mathrm{esc, Hf}} \sim 1.
\end{equation}

There is effectively no change which is consistent with the similar effective transmission between the two devices. With the higher gap energy in the bilayer, these results suggest that the bilayer should have worse resolving power than that of the hafnium device. Since this is not the case, we infer that an alternative mechanism must be preventing phonon transmission. In appendix~\ref{sec:improvement}, the size of the phonon trapping effect caused by the lack of high energy phonon states in indium is estimated to reduce $J$ by between 1.7 and 26 which is consistent with our measurements. The uncertainty in this estimation is dominated by the unknown material constants for hafnium.

In conclusion, we have fabricated an optical MKID made out of an indium / hafnium bilayer. We find that the resolving power of this device is nearly twice what has been previously measured in other single layer devices and approaches the best resolving powers measured in membrane suspended MKIDs by achieving a similar amount of phonon trapping. Simulations of the phonon propagation across the extra interface layer do not adequately explain the improved resolving power and order of magnitude estimates of the energy down-conversion physics point to the low phonon cutoff energy in indium as the primary phonon trapping mechanism. These results show that the simple addition of an extra layer to the MKID sensor material can significantly improve the detector performance, approaching the minimum requirements for an effective exoplanet IFS. Additionally, while the motivation of this letter was focused on exoplanet instrumentation, this technique may reduce the fabrication complexity of detectors where $R\sim20$ at 1 $\mu$m is acceptable, like for bio-analysis research~\cite{Niwa2017} or for dark matter detection~\cite{Baryakhtar2018}. Further improvements to the resolving power in optical MKIDs may involve testing different interface materials or combining the membrane suspension and phonon trapping layer techniques.

The low Debye energy interface layer demonstrated here is an interesting tool for manipulating phonon dynamics at low temperatures and may also have uses in other types of devices. As shown in figure~\ref{fig:diagram}, it acts as a selective valve, preventing highly non-equilibrium phonons from crossing the barrier, while allowing passage for low energy phonons which help keep the device in thermal equilibrium with the substrate. We use it here to keep high energy phonons inside of our detector, but the interface layer should also provide protection to the device from absorbing high energy phonons generated in the substrate. One potential source of these phonons is from ionizing radiation, like cosmic rays, which cause detector glitches and lead to a significant increase in dead time in superconducting bolometers~\cite{Karatsu2019}. These types of events also pose problems for quantum computers by destroying qubit coherence and increasing error rates~\cite{Vepsalainen2020,Martinis2021,McEwen2022}. This issue may be partially mitigated by implementing a similar interface layer in these respective devices.

\begin{acknowledgments}
\section{Acknowledgments}
N.Z. was supported throughout this work by a NASA Space
Technology Research Fellowship. This material is based upon work supported by the National Aeronautics and Space Administration under grant number 80NSSC19K0329.
\end{acknowledgments}

\appendix

\counterwithout{equation}{section}
\renewcommand{\theequation}{S\arabic{equation}}
\renewcommand{\thefigure}{S\arabic{figure}}  
\renewcommand{\thetable}{S\Roman{table}} 
\setcounter{secnumdepth}{2}

\section{Resolving Power Measurement} \label{sec:r}
Our MKID readout is a traditional homodyne readout scheme using a kinetic inductance traveling wave amplifier operating in the three-wave mixing mode. This experimental setup is described in reference~\onlinecite{Zobrist2019}.

We first characterize the forward scattering parameter, $S_{21}$, for our system by sweeping a range of frequencies surrounding the resonance and measuring the resonance circle at a variety of input powers. We pick a bias power 1 to 2 dB lower than the saturation point of the resonator to increase the signal to noise while avoiding any hysteretic switching from the circle bifurcation~\cite{Swenson2013}. The probe tone is set to the resonance frequency to maximize the detector response.

Lasers at different energies are used to determine the detector response and resolving power. Before turning on a laser, 10 seconds of noise data is captured. Each laser is then turned on one at a time, and 10,000 photon absorption events are recorded per laser energy. The count rate is kept below 100 photons per second to prevent pulse pileup effects.  

To determine the resolving power, the I and Q signals from the mixer are converted to the phase and dissipation coordinates derived in reference~\onlinecite{Zobrist2020}, which increase the dynamic range of the detector. For each laser energy, a template is constructed from the average detector response and combined with the noise power spectrum into a filter. Under the assumptions of detector linearity, additive Gaussian noise, and low count rates, this type of filter results in the lowest variance estimate of the photon energies when convolved with the data~\cite{Fowler2016}.

\section{Effective phonon transmission} \label{sec:transmission}
We use the average phonon transmission from material $i$ to adjacent material $j$ defined by the appropriate combination of the transmission coefficients for longitudinal and transverse phonons~\cite{Kaplan1979}.
\begin{equation}
    T_{ij} = \left(\frac{2 T_{t, ij}}{3 c_{t, i}^2} + \frac{T_{l, ij}}{3 c_{l, i}^2} \right)\left(\frac{2}{3 c_{t, i}^3} + \frac{1}{3c_{l, i}^3}\right)^{-2/3}
\end{equation}
Here $c$ corresponds to the speed of sound and the $t$ and $l$ subscripts specify the corresponding transverse or longitudinal phonon mode. The average reflection coefficient can be found with $R_{ij} = 1 - T_{ij}$.

To compute the effective transmission from material~1, through material~2, and into material~3, we add up all of the transmission and reflection paths that result in a phonon entering material~3, assuming no phonon coherence across the structure. This procedure results in the following formula:
\begin{equation} \label{eq:t13}
    T_{13} = T_{12} T_{23} \sum^\infty_{i=0} \left(R_{23} R_{21} \right)^i.
\end{equation}

Table~\ref{tab:materials} contains all of the material parameters used to compute the transmission coefficients from the acoustic mismatch model and the effective transmission through an interface defined by equation~\ref{eq:t13}. 
\begin{table}[h]
    \centering
    \begin{tabularx}{\linewidth}{cCCC}
    \toprule
     Material \phantom{a} & density [\si{\g \per \cm^3}] & $c_t$ [\si{\m \per \s}] & $c_l$ [\si{\m \per \s}] \\
     \midrule
     Hafnium & 12.781 & 2053 & 3786 \\
     Indium & 7.47 & 904 & 2700 \\
     Silicon & 2.33 & 5340 & 8980 \\
     Sapphire & 3.99 & 6450 & 10900 \\
     PMMA & 1.18 & 1400 & 2757 \\
     \bottomrule
    \end{tabularx}
    \caption{Material parameters used in the acoustic mismatch calculations. All data were taken from reference~\onlinecite{Kaplan1979} with the exception of those for hafnium and PMMA which were found in reference~\onlinecite{Qi2016} and~\onlinecite{Destgeer2017} respectively and correspond to room temperature measurements.}
    \label{tab:materials}
\end{table}

\section{Fabrication Details} \label{sec:fab}

\begin{table*}[t]
\centering
\begin{tabularx}{\linewidth}{CCCCCCC}
\toprule
 Material &  d [nm] & $\rho$ @ 4 K [\si{\micro\ohm \cm}] & $\Theta_D$ [K] & $T_c$ [K] & $N_0$ [\si{\per \joule \per \m \cubed}] & $D$ [\si{\cm \squared \per \s}] \\
 \midrule
 Hafnium & 220 & 60.6 & 252 & 0.468 & $1.23 \times 10^{47}$ & 5.24 \\
 Indium & 15 & 0.0973 & 112 & 3.3 & $8.78 \times 10^{46}$ & 4570 \\
 \bottomrule
\end{tabularx}
\caption{The layer thicknesses, resistivities, Debye temperatures, transition temperatures, single-spin density of states, and diffusion constants used in the numerical simulation are tabulated here. The single-spin density of states includes the phonon enhancement factor and was computed using the data from reference~\onlinecite{McMillan1968} and the densities presented in table~\ref{tab:materials}. The diffusion constant was computed using $D = \sfrac{1}{\rho N_0 e^2}$~\cite{Martinis2000}. All other parameters were measured directly from films deposited on silicon wafers with deposition parameters matching those used to make the bilayer.}
\label{tab:dos}
\end{table*}

The silicon wafer is mounted into the vacuum chamber of a thermal evaporator. Indium drops are placed into a tungsten crucible and the system is pumped down to a base pressure of $3 \times 10^{-6}$ Torr where 15 nm of indium is evaporated onto the substrate. 

The substrate is then transferred into a ultra high vacuum AJA ATC-2200 sputter deposition chamber and 220 nm of hafnium are deposited on top of the indium. The hafnium deposition process is described fully in reference \onlinecite{Coiffard2020}, but we use a different sputter target. The target is \mbox{$>$99.95\%} hafnium by weight of the 79 elements sampled excluding zirconium. Zr, O, C, N, Ta, Fe, H, Nb, Al, Si are all 700, 60, 10, \mbox{$<$10}, \mbox{$<$5}, 1.5, \mbox{$<$0.5}, 0.25, 0.24, 0.21~ppm-wt respectively. All other elements are \mbox{$\le$0.14ppm-wt}.  

The wafer then goes through a single step of lithography consisting of the definition of the MKID resonators. 80 nm of DUV-42P6 adhesion promoter is spun on the wafer followed by 800 nm of imaging UV6-0.8 photoresist. The MKID geometry is patterned with the same stepper and the indium/hafnium bilayer is etched in a PlasmaTherm SLT 700 reactive ion etcher in a BCL$_3$/Cl$_2$ environment (60/40 sccm, 5 mT, 100 W). Finally, the resist is removed with solvents and gold bond pads are added via a lift off process on the side of the chip to ensure good thermalization of the device. The wafer is then diced into chips of dimension 13x13 mm.

Between the deposition of indium and hafnium, the sample is exposed to atmosphere. A small uncharacterized oxide layer may form which is not considered in the phonon transport calculations in this letter. Future work will investigate if removing this layer has a significant impact on the detector resolving power.

\section{Bilayer Proximity Effect and $T_c$} \label{sec:proximity}
We measure $T_c = \SI{468}{\milli \K}$ for hafnium on silicon and \SI{786}{\milli \K} for the bilayer on silicon indicating that the thin indium layer has a strong effect on the superconducting properties of the film. Luckily, \SI{786}{\milli \K} is still within the sub-Kelvin value range needed for this type of detector, so we should not be too concerned. At most, the detector thickness will need to be tuned in the future to compensate for the slight decrease in responsivity. As the hafnium is already fairly thick, we can be confident that this issue will not constrain detector performance.

\begin{figure}[t]
    \centering
    \includegraphics[width=0.5\textwidth]{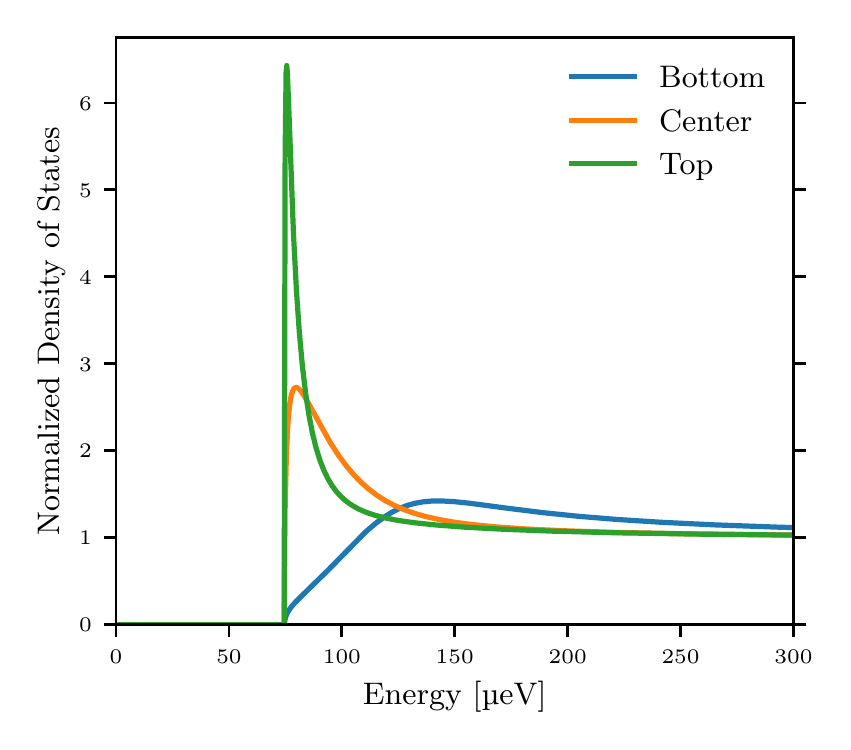}
    \caption{The normalized density of states for three vertical positions along the bilayer are plotted. \textit{Top} corresponds to the side of the film furthest from the substrate while \textit{Bottom} refers to the interface between the indium and hafnium. The density of states of the hafnium near the interface and all of the indium is indistinguishable from the \textit{Bottom} curve. The density of states in the center of the hafnium film is also shown for comparison.}
    \label{fig:dos}
\end{figure}

To try to explain this variation in $T_c$ we calculate how the density of states in the film changes due to the proximity effect. We use a numerical simulation based on the Usadel equations to calculate the density of states as a function of vertical position in the film to verify that it has a single gap energy~\cite{Zhao2018a}. Figure \ref{fig:dos} shows the results of this computation using the parameters defined in table \ref{tab:dos} and assuming zero boundary resistance. The film does have a single gap energy despite significant variation in the density of states. The predicted variation in the density of states does not effect detector performance because the size of the diffusion constant in each material results in a uniform vertical distribution of quasiparticles on \si{\micro \s} timescales for this detector thickness. 

However, the calculation shows that the film transition temperature should only increase by a few \si{\milli \K} when adding the thin indium layer which does not match our measurements. The real film may have a significant boundary resistance due to an indium oxide layer, but reasonable values for this resistance fail to make up for the difference between measured and calculated $T_c$. This is somewhat expected because the hafnium $T_c$ depends strongly on the film stress and substrate used~\cite{Coiffard2020}. We should expect, then, that the bilayer $T_c$ is similarly controlled by the deposition parameters of the film and less so by the proximity effect's influence.

\section{Quasiparticle Lifetime} \label{sec:qp}
\begin{figure*}[t]
    \centering
    \includegraphics[width=0.994\textwidth]{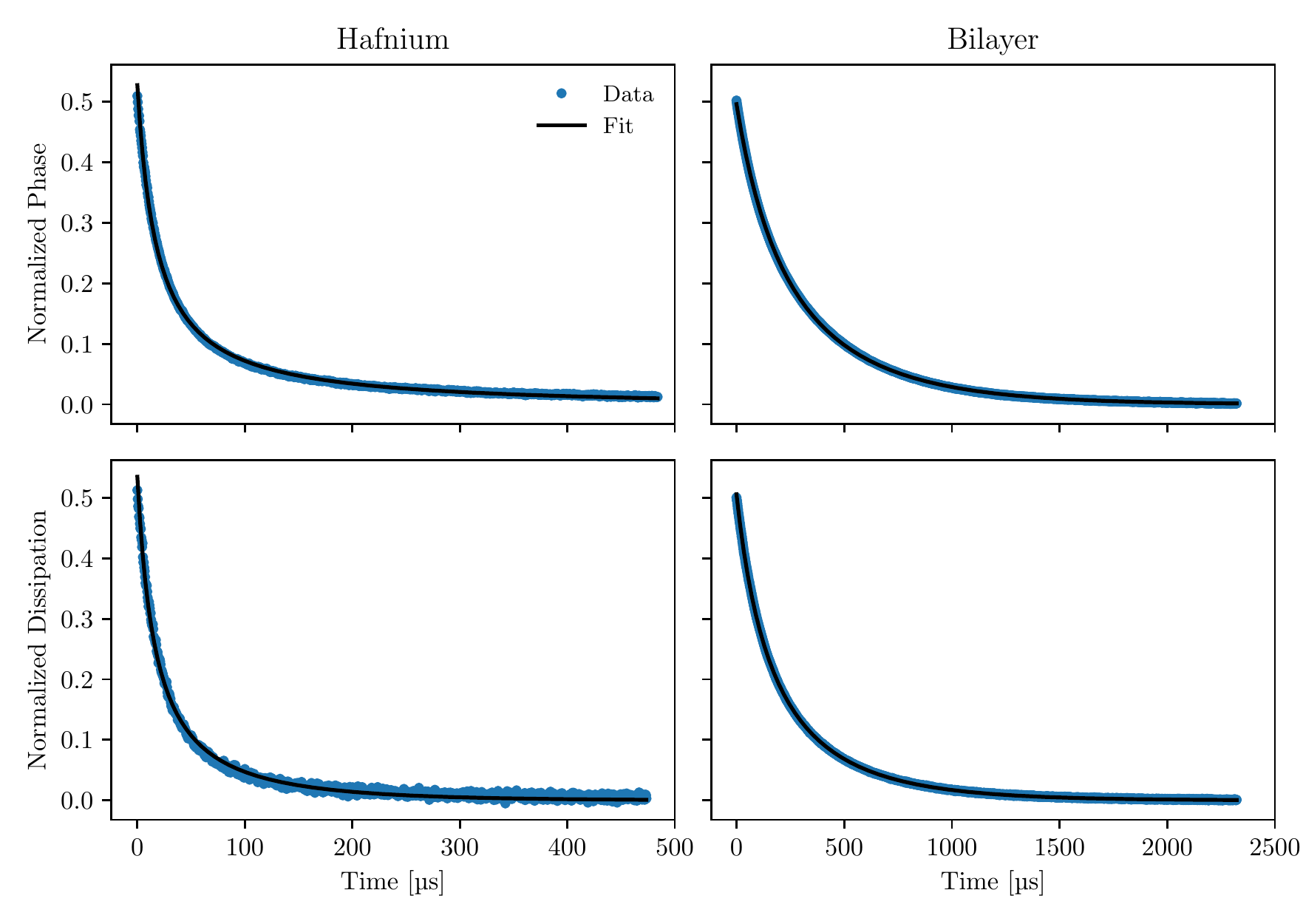}
    \caption{Fits to the average pulse decay in the phase and dissipation quadratures are shown for both the hafnium film and hafnium bilayer. Each pulse response is normalized to its pulse height. }
    \label{fig:qp}
\end{figure*}

It can be difficult to extract the quasiparticle lifetime in a superconductor based on a photon response. The quasiparticle recombination rate is quadratic in the total number of quasiparticles, which results in a pulse shape that transitions from a hyperbolic to exponential decay over time. The functional form for this decay was worked out in reference~\onlinecite{Fyhrie2018}.
\begin{equation}
    X(t) = \frac{A}{\left[ 1 + \sfrac{1}{x_{qp}(0)} \right]\mathrm{exp}\left(\sfrac{t}{\tau_{qp}} \right) - 1}
\end{equation}
Here, $X$ is either the phase or dissipation response. $t$ is time. $A$ is a scaling parameter. $x_{qp}(0)$ is the fractional quasiparticle density at the beginning of the pulse. $\tau_{qp}$ is the quasiparticle lifetime at zero quasiparticle density. 

$\tau_{qp}$ differs from the pulse decay time presented for hafnium in reference~\onlinecite{Zobrist2019a}, for example, because it is not a time constant representative of the pulse as a whole but rather of the tail end of the pulse. Figure~\ref{fig:qp} shows the fits to the average pulse in each quadrature of the signal to this model. We choose to limit the fit to data less than half of the pulse peak to minimize the effect of the finite resonator ring up time on the results. The fit parameters and their statistical uncertainties are given in table~\ref{tab:qp}.

\begin{table}[h]
\centering
\begin{tabularx}{\linewidth}{CCCC}
\toprule
 Film & Response &  $\tau_{qp}$ & $x_{qp}$ \\
 \midrule
 \multirow{2}{*}{Hafnium} & Phase & $404 \pm 6.3$ & $21.6 \pm 0.42$  \\
 & Dissipation & $106 \pm 1.7$ & $5.6 \pm 0.15$ \\
 \midrule
 \multirow{2}{*}{Bilayer} & Phase & $452.4 \pm 0.55$ & $1.010 \pm 0.0044$  \\
 & Dissipation & $401.3 \pm 0.48$ & $1.585 \pm 0.0053$ \\
 \bottomrule
\end{tabularx}
\caption{The fitted values for the quasiparticle lifetime and initial quasiparticle fractional density along with their 1-$\sigma$ uncertainties are tabulated for the phase and dissipation directions. }
\label{tab:qp}
\end{table}

\section{Phonon Escape Time Simulation} \label{sec:sim}
We employ a similar model as discussed in reference~\onlinecite{deVisser2021} to calculate the phonon escape time, $\tau_\mathrm{esc}$. The geometry is assumed to be a 2 $\mu$m wide rectangle with infinite length. The hafnium thickness is 220 nm and the indium thickness is 15 nm. Phonons are emitted in random directions from the top of the film with equal probability for each phonon mode. When a phonon hits an interface it is either transmitted at a new direction determined by Snell's law or reflected.  If reflected, diffuse scattering is assumed and a random direction is chosen. Care is taken at the interfaces to calculate the probability of mode conversion between transverse and longitudinal phonons according to the acoustic mismatch model~\cite{Kaplan1979}. $\tau_\mathrm{esc}$ is computed by keeping track of the time required for each phonon to enter the substrate and averaging that time for 10,000 different starting conditions.

\section{Expected Phonon Trapping Improvement} \label{sec:improvement}
Sputter deposited thin film hafnium differs significantly from bulk hafnium. The thin film transition temperature is a factor of 4 higher than its bulk value~\cite{Coiffard2020}. It also is a very disordered material resulting in an anomalously high normal state resistivity~\cite{Kraft1998}. With this in mind, it is unsurprising that the material constants for hafnium provided in reference~\onlinecite{Kozorezov2000} do not provide sensible answers when calculating the phonon loss factor, $J$. We do, however, expect that the general picture of the energy down-conversion and phonon loss still apply~\cite{Kozorezov2007, Kozorezov2008}. Using the formulas from these references, we can put bounds on our expected change in $J$ when the hafnium bilayer is introduced. 

The phonon loss can be partitioned into two contributions, $J=J_\mathrm{high} + J_\mathrm{low}$. $J_\mathrm{high}$ represents the initial wave of phonons created as the hot electron plasma cools to the Debye temperature of the superconductor. There are relatively few of these phonons and they each carry a significant fraction of the original photon's energy. The formula for $J_\mathrm{high}$ is fairly complicated, but since we are only looking at phonon energy scaling we can ignore any constant or weakly energy dependent terms~\cite{Kozorezov2007}.
\begin{equation}
    J_\mathrm{high} \propto \int^{\Omega_D}_0 d\epsilon \; \epsilon^4 l_\text{pb}(\epsilon)
\end{equation}
The integral sums up the contributions to the energy uncertainty from phonons with energies between 0 and the Debye energy, $\Omega_D$. The phonon mean-free-path due to pair breaking, $l_\text{pb}(\epsilon)$, scales inversely with energy, making the integrand scale as $\epsilon^3$~\cite{Kaplan1976}.

Introducing an interface layer would lower the upper bound on this integral to the Debye energy of the interface. However, all other references to the Debye energy in the equation for $J_\text{high}$ should remain unchanged since the energy scale of the phonon distribution is set by the acoustic properties of the hafnium film. Taking the Debye temperature of indium to be 112 K and the Debye temperature of hafnium to be 252 K~\cite{McMillan1968}, we find
\begin{equation}
    \frac{J^\mathrm{Hf}_\mathrm{high}}{J^\mathrm{Hf, In}_\mathrm{high}} = 26.
\end{equation}
We expect a significant reduction in the amount of energy loss from this first generation of phonons. We should note that because of the power of 4 energy scaling, these results are very sensitive to the exact values used for the Debye energies.

The effect of successive generations of phonons are more difficult to account for. The energy dependence of $J_\mathrm{low}$ is contained in the function $g_1\left(\sfrac{\Omega_D}{\Omega_1}\right)$ presented in reference~\onlinecite{Kozorezov2008} and corrected in reference~\onlinecite{deVisser2021}. $\Omega_1$ is the transition energy from a phonon dominated down-conversion to a quasiparticle dominated down-conversion. Therefore, the energy scaling can not be disentangled from $\Omega_1$. Using the $\Omega_1$ given in reference~\onlinecite{Kozorezov2000}, we find
\begin{equation}
    \frac{J^\mathrm{Hf}_\mathrm{low}}{J^\mathrm{Hf, In}_\mathrm{low}} = 1.7.
\end{equation}
This value is consistent with less lower energy phonons being blocked by the difference in Debye temperature.

Since we do not know the relative contributions of $J_\mathrm{high}$ and $J_\mathrm{low}$, these two values put bounds on the improvement to the total phonon loss factor, which is consistent with the measured improvement of 8. 
\begin{equation}
    1.7<\frac{J^\mathrm{Hf}}{J^\mathrm{Hf, In}} < 26.
\end{equation}
This model is consistent with our data. However, more work needs to be done to understand the exact phonon material parameters for these films. 

\bibliography{references}

\end{document}